\documentclass[aps,prd,nofootinbib,amsmath,amssymb,showpacs,superscriptaddress,twocolumn]{revtex4}
\usepackage{txfonts}
\usepackage{graphicx}
\usepackage{dcolumn}
\usepackage{bm}
\usepackage{amssymb}
\usepackage{latexsym}
\usepackage[colorlinks, linkcolor=blue, citecolor=blue, urlcolor=blue]{hyperref}

\newcommand{\be}{\begin{equation}}
\newcommand{\ee}{\end{equation}}

\def\bea{\begin{eqnarray}}
\def\eea{\end{eqnarray}}
\begin{document}

\title{Constraining dynamical dark energy with a divergence-free parametrization in the presence of spatial curvature and massive neutrinos}

\author{Hong Li}
\affiliation{Institute of High Energy Physics, Chinese Academy of
Sciences, Beijing 100049, China} \affiliation{National Astronomical
Observatories, Chinese Academy of Sciences, Beijing 100012, China}
\author{Xin Zhang\footnote{Corresponding author}}
\email{zhangxin@mail.neu.edu.cn} \affiliation{Department of Physics,
College of Sciences, Northeastern University, Shenyang 110004,
China} \affiliation{Center for High Energy Physics, Peking
University, Beijing 100080, China}

\begin{abstract}
In this paper, we report the results of constraining the dynamical
dark energy with a divergence-free parameterization, $w(z) = w_{0} +
w_{a}\left(\frac{\ln(2+z)}{1+z}-\ln2\right)$, in the presence of
spatial curvature and massive neutrinos, with the 7-yr WMAP
temperature and polarization data, the power spectrum of LRGs
derived from SDSS DR7, the Type Ia supernova data from Union2
sample, and the new measurements of $H_0$ from HST, by using a MCMC
global fit method. Our focus is on the determinations of the spatial
curvature, $\Omega_k$, and the total mass of neutrinos, $\sum
m_{\nu}$, in such a dynamical dark energy scenario, and the
influence of these factors to the constraints on the dark energy
parameters, $w_0$ and $w_a$. We show that $\Omega_k$ and $\sum
m_{\nu}$ can be well constrained in this model; the $95\%$ CL limits
are: $-0.0153<\Omega_k<0.0167$ and $\sum m_{\nu}<0.56$ eV. Comparing
to the case in a flat universe, we find that the error in $w_0$ is
amplified by $25.51\%$, and the error in $w_a$ is amplified by
$0.14\%$; comparing to the case with a zero neutrino mass, we find
that the error in $w_0$ is amplified by $12.24\%$, and the error in
$w_a$ is amplified by $1.63\%$.
\end{abstract}



\maketitle


\section{Introduction}\label{sec:intro}

Dark energy is one of the most important themes in physics today.
However, we do not know much about dark energy due to the accuracy
of current data. Though the current observational data are
consistent with a cosmological constant, the possibility that dark
energy is dynamical is still not excluded by the data and has been
attracting wide attentions in the cosmology and theoretical physics
communities.

In order to detect the dynamics of dark energy, one usually has to
parameterize the equation-of-state parameter (EOS), $w$,
empirically, using two or more free parameters. Among all the
parametrization forms of EOS, the Chevallier-Polarski-Linder (CPL)
parametrization~\cite{CPL}, $w(a)=w_0+w_a(1-a)$, where $w_0$ and
$w_a$ are parameters and $a$ is the scale factor of the universe, is
the most widely used one and has been explored extensively. However,
as pointed out in Ref.~\cite{Ma:2011nc}, the CPL description will
lead to unrealistic behavior in the future evolution, i.e., $|w(z)|$
grows rapidly and eventually encounters divergence as the redshift
$z$ approaches $-1$. In order to keep the advantage of the CPL
parametrization and avoid its drawback at the same time, it is
believed that a parametrization that is free of divergence both in
the past and future evolutions is necessary.

In Ref.~\cite{Ma:2011nc}, Ma and Zhang proposed the following hybrid
form of logarithm and CPL parametrizations:
\begin{equation}\label{MZ}
w(z) = w_{0} + w_{a}\left(\frac{\ln(2+z)}{1+z}-\ln2\right).
\end{equation}
This novel parametrization has well behaved, bounded evolution for
both high redshifts and negative redshifts. In particular, for the
limiting situation, $z\rightarrow -1$, a finite value for EOS can be
obtained, $w=w_0+w_a(1-\ln 2)$. At low redshifts, this
parametrization form reduces to the linear one, $w(z)\approx
w_0+\tilde{w}_az$, where $\tilde{w}_a=-(\ln2)w_a$. Of course, one
can also recast it at low redshifts as the CPL form, $w(z)\approx
w_0+\tilde{w}_az/(1+z)$, where $\tilde{w}_a=(1/2-\ln2)w_a$.
Therefore, it is clear to see that this parametrization exhibits
well-behaved feature for the dynamical evolution of dark energy.
Without question, such a two-parameter form of EOS can genuinely
cover scalar-field models as well as other theoretical scenarios. In
Ref.~\cite{fate}, this parametrization was used to explore the
ultimate fate of the universe.

The parametrization (\ref{MZ}) has been explored deeply by using the
current data. In particular, in Ref.~\cite{Li:2011dr}, we have
analyzed the detection of dynamics of dark energy with the
parametrization (\ref{MZ}) by performing a Markov Chain Monte Carlo
(MCMC) global fitting method. Since dark energy parameters are
tightly correlated with the spatial curvature $\Omega_k$ and
neutrino mass $\sum m_{\nu}$, in this paper we will deeply analyze
the influences of these factors on the detection of dynamics of dark
energy.

The paper is organized as follows: In Sec.~\ref{sec:data} we will
introduce our global fitting procedure and the data we used for
analysis; the results are presented in Sec.~\ref{sec:result}, and
our conclusion is given in Sec.~\ref{sec:conclu}.

\section{Global fitting procedure and Data}\label{sec:data}

We have modified MCMC package CosmoMC~\cite{CosmoMC} to perform a
global fitting analysis for the dynamical dark energy parameterized
above. As we know, within the dynamical dark energy models, the
perturbations of dark energy are
important~\cite{WMAP3GF,LewisPert,XiaPert} for the data fitting
analysis. For quintessencelike or phantomlike models, whose $w$ does
not cross the cosmological constant boundary, the perturbation of
dark energy is well defined. However, when $w$ crosses $-1$, which
is described by a quintom dark energy model~\cite{Feng:2004ff}, one
is encountered with the divergence problem for perturbations of dark
energy at $w=-1$. For avoiding such kind of divergence problem, in
this paper we use the method provided in
Refs.~\cite{XiaPert,Zhao:2005vj} to treat the dark energy
perturbations consistently in the whole parameter space in the
numerical calculations.

Our most general parameter space vector is:
\begin{equation}
\label{parameter} {\bf P} \equiv (\omega_{b}, \omega_{c}, \Theta,
\tau, w_{0}, w_{a}, \Omega_k, \sum m_{\nu}, n_{s}, A_{s}, c_s^2),
\end{equation}
where $\omega_{b}\equiv\Omega_{b}h^{2}$ and
$\omega_{c}\equiv\Omega_{c}h^{2}$, with $\Omega_{b}$ and
$\Omega_{c}$ the physical baryon and cold dark matter densities
relative to the critical density, $\Omega_k$ is the spatial
curvature satisfying $\Omega_k+\Omega_m+\Omega_{\rm de}=1$, $\sum
m_{\nu}$ is the total mass of neutrinos, $\Theta$ is the ratio
(multiplied by 100) of the sound horizon to the angular diameter
distance at decoupling, $\tau$ is the optical depth to
re-ionization, $w_0$ and $w_a$ are the parameters of dark energy EOS
given by Eq.~(\ref{MZ}), $A_s$ and $n_s$ are the amplitude and the
spectral index of the primordial scalar perturbation power spectrum,
and $c_s$ is the sound speed of dark energy. For the pivot scale we
set $k_{s0}=0.05$Mpc$^{-1}$. Note that we have assumed purely
adiabatic initial conditions.

Note that the sound speed of dark energy is fixed in our analysis.
In the framework of the linear perturbation theory, besides the EOS
of dark energy, the dark energy perturbations can also be
characterized by the sound speed, $c_s^2\equiv\delta p_{\rm
de}/\delta\rho_{\rm de}$. The sound speed of dark energy might
affect the evolution of perturbations, and might leave signatures on
the CMB power spectrum \cite{Xia:2007km}. However, it has been shown
that the constraints on the dark energy sound speed $c_s^2$ in
dynamical dark energy models are still very weak, since the current
observational data are still not accurate enough \cite{Li:2010ac}.
Therefore, in our analysis, we have treated the dark energy as a
scalar-field model (multi-fields or single field with high
derivative) and set $c_s^2$ to be 1. Of course, one can also take
$c_s^2$ as a parameter, but the fit results would not be affected by
this treatment \cite{Li:2010ac}.

In the computation of the cosmic microwave background anisotropy
(CMB), we include the 7-year WMAP temperature and polarization power
spectra \cite{Komatsu:2010fb} with the routine for computing the
likelihood supplied by the WMAP team~\cite{code}. For the large
scale structure (LSS) information, we use the power spectrum of
luminous red galaxies (LRGs) measured from the SDSS DR7
\cite{Abazajian:2008wr}. The supernova (SN) data we use are the
recently released ``Union2'' sample of 557 data \cite{Riess:2011yx};
note that the systematic errors are included in our analysis. In the
calculation of the likelihood from SN we marginalize over the
relevant nuisance parameter \cite{SNMethod}.

Furthermore, we make use of the Hubble Space Telescope (HST)
measurement of the Hubble constant $H_{0}\equiv 100
h$~km~s$^{-1}$~Mpc$^{-1}$ by a Gaussian likelihood function centered
around $h=0.738$ and with a standard deviation $\sigma=0.024$
\cite{Riess:2011yx}.

\section{Numerical Results}\label{sec:result}

In this section we shall present the results of the global fitting
analysis. In particular, we shall test the influences on probing the
dynamics of dark energy when spatial curvature and massive neutrinos
are considered in the data analysis. Since we know that both the
spatial curvature parameter $\Omega_k$ and the total mass of
neutrinos $\sum m_{\nu}$ are degenerate with the equation of state
of dark energy $w$, we are very much interested in seeing how these
two factors affect the constraints on the parameters $w_0$ and
$w_a$. Moreover, we also want to see how the current data constrain
$\Omega_k$ and $\sum m_{\nu}$ within the framework of dynamical dark
energy with parametrization (\ref{MZ}).

\subsection{Spatial curvature and EOS}\label{subsec:curvature}

Dark energy parameters and $\Omega_k$ are correlated via the
cosmological distance information. From the observation of CMB, the
curvature of the observable universe $\Omega_k$ can be determined by
the position of first acoustic peak of CMB temperature power
spectrum precisely. However, $\Omega_k$ is tightly degenerated with
$\Omega_m$, and such degeneracy can be, in certain level, broken by
taking into account the data of large scale structure and
supernovae.

In constraining $\Omega_k$, one usually uses a combination of
distance measurements from baryon acoustic oscillations (BAO) and
CMB. In this paper, however, we use the power spectrum of LRGs
measured from SDSS DR7, instead of BAO, to constrain $\Omega_k$.
Note that BAO and the LRG power spectrum cannot be treated as
independent data sets because a part of the measurement of BAO comes
from LRGs as well. Since the LRG power spectrum is a powerful probe
of the total mass of neutrinos, $\sum m_{\nu}$, in this paper we
uniformly use the LRG power spectrum, instead of BAO, to constrain
$\Omega_k$ and $\sum m_{\nu}$. In the following we shall consider
the case of massless neutrinos and see the influence of $\Omega_k$.

\begin{figure}[t]
\begin{center}
\includegraphics[scale=0.5]{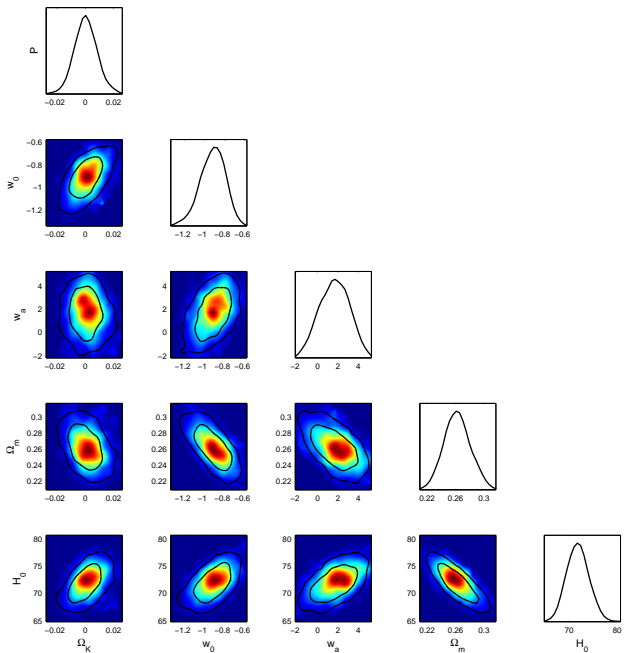}
\caption{Constraints on the cosmological parameters $\Omega_k$,
$w_0$, $w_a$, $\Omega_m$ and $H_0$ in the dynamical dark energy
scenario (\ref{MZ}).\label{fig1}}
\end{center}
\end{figure}

In Fig.~\ref{fig1}, we present the two dimensional cross correlation
and one dimensional probability distribution of $\Omega_k$, $w_0$,
$w_a$, $\Omega_m$ and $H_0$ by fitting with 7-yr WMAP, LRG, SN and
$H_0$ data. From the panels of $\Omega_k$--$w_0$ and
$\Omega_k$--$w_a$, we find that the correlation between $\Omega_k$
and $w_0$ is positive, but we do not observe much correlation
between $\Omega_k$ and $w_a$.

The $68\%$ confidence level (CL) constraint on $\Omega_k$ is $0.0003
\pm0.0079$, from which we can see that our universe is very close to
flatness. The $95\%$ CL limit is: $-0.0153<\Omega_k<0.0167$. So, we
find that in the dynamical dark energy model (\ref{MZ}) the spatial
curvature is well constrained by WMAP+LRG+SN+$H_0$. It is of
interest to make a comparison to the cases of $\Lambda$CDM ($w=-1$)
model and constant $w$ model. Komatsu et al. gave the limit on
$\Omega_k$ for the case of $w=-1$,
$\Omega_k=-0.0023^{+0.0054}_{-0.0056}$ ($68\%$ CL), from
WMAP+BAO+$H_0$, and for the case of constant $w$,
$\Omega_k=-0.0057^{+0.0066}_{-0.0068}$ ($68\%$ CL), from
WMAP+BAO+SN, where SN is the Constitution sample. These results are
in good agreement with our limit.

How does the spatial curvature parameter, $\Omega_k$, affect the
constraint results of ($w_0$, $w_a$)? We find that, in a non-flat
universe, the results are:
$$w_0=-0.922\pm0.123~ {\rm and}~ w_a=1.651\pm1.470~{\rm (68\%~ CL)}.$$
Thus, even when $w$ is allowed to depend on time, the current data
are still consistent with a cosmological constant. However, a large
range of values of ($w_0$, $w_a$) are still allowed by the data.
Comparing to the case in a flat universe: $w_0=-0.921\pm0.098$ and
$w_a=1.905\pm1.468$ ($68\%$ CL), we find that the best-fit values of
$w_0$ and $w_a$ are shifted by $0.1\%$ and $13.3\%$, respectively;
the error in $w_0$ is amplified by $25.51\%$, and the error in $w_a$
is almost the same, only amplified by $0.14\%$.

One may be curious to know why the error on $w_0$ is significantly
amplified, but that for $w_a$ is scarcely affected by the presence
of $\Omega_k$. Actually, this is fairly easy to be understood. From
the fitting results, it is clear to see that $w_0$ is constrained
more tightly than $w_a$, i.e., $w_0$ is more sensitive to the data,
comparing to $w_a$. The current constraint on $w_a$ is very weak,
only at the level of $\sim 80\%$, while for $w_0$ the constraint is
much stronger, at the level of $\sim 10\%$. Thus, once $\Omega_k$ is
involved in the analysis, its influence on $w_0$ is, obviously, much
greater than that on $w_a$.

Of course, the fact that $w_a$ is insensitive to the data is not
specific to our parametrization form. In general, for a
two-parameter dark energy EOS form in which $w_0$ serves as a
constant part and $w_a$ is used to describe the dynamical evolution
of the EOS, $w_a$ is always insensitive to the data comparing to
$w_0$. The constraints on $w_0$ and $w_a$ are mainly determined from
their contributions to the Hubble expansion rate $H(z)$ which is
more sensitive to $w_0$ than $w_a$. During the calculation,
considering the same variations of $w_0$ and $w_a$, the effect on
$H(z)$ from $w_0$ is more significant than that from $w_a$. Thus,
when using the observational data to explore dark energy EOS, the
constraint on $w_0$ is usually better than that on $w_a$ by about
one order of magnitude. This explains the fact that the
time-evolution of dark energy EOS cannot be well probed by the
current cosmological data. Taking the number of parameters into
account, it has been shown that the $\Lambda$CDM model is still the
best one among all the current dark energy models; see, e.g.,
Refs.~\cite{Davis:2007na,Li:2009jx}. It is expected that the dark
energy observations of the next generation could measure the
dynamical evolution of dark energy accurately.

So, our conclusion that $w_0$ is affected by $\Omega_k$ more
severely than $w_a$ is universal, not specific to the
parametrization adopted in this paper. Obviously, the same
conclusion is applicable to the case of massive neutrinos that will
be discussed in the next subsection.

\subsection{Neutrino mass and EOS}\label{subsec:neutrino}

Weighting the neutrino mass is one of the most important challenges
in modern physics. Currently the neutrino oscillation experiments,
such as atmospheric neutrinos experiments \cite{Atmospheric} and
solar neutrinos experiments \cite{Solar}, have confirmed that the
neutrinos are massive, but give no hint on their absolute mass
scale. Cosmology can provide crucial complementary information on
absolute neutrino masses, because massive neutrinos leave imprints
on the cosmological observations, such as the Hubble diagram, CMB
temperature power spectrum and LSS matter power spectrum
\cite{NeuRev}.

In Ref.~\cite{Komatsu:2010fb}, Komatsu et al. have constrained the
total mass of neutrinos, $\sum m_{\nu}=94{\rm eV}(\Omega_{\nu}h^2)$,
from the 7-yr WMAP data combined with the distance information. For
a flat $\Lambda$CDM model, i.e., $w=-1$ and $\Omega_k=0$, they found
that the WMAP+BAO+$H_0$ limit is $\sum m_{\nu}<0.58$ eV ($95\%$ CL).
The limit improves when information on the growth of structure is
added. For example, they found that, when the BAO is replaced by the
power spectrum of LRGs, the combination WMAP+LRG+$H_0$ gives $\sum
m_{\nu}<0.44$ eV ($95\%$ CL) for $w=-1$. Thus, we can see that the
power spectrum of LRGs plays an important role in constraining the
neutrino mass. In the following we shall use the combination of
WMAP+LRG+SN+$H_0$ to constrain the neutrino mass, $\sum m_{\nu}$, as
well as the parameters of dynamical dark energy, $w_0$ and $w_a$, in
a flat universe.

\begin{figure}[t]
\begin{center}
\includegraphics[scale=0.5]{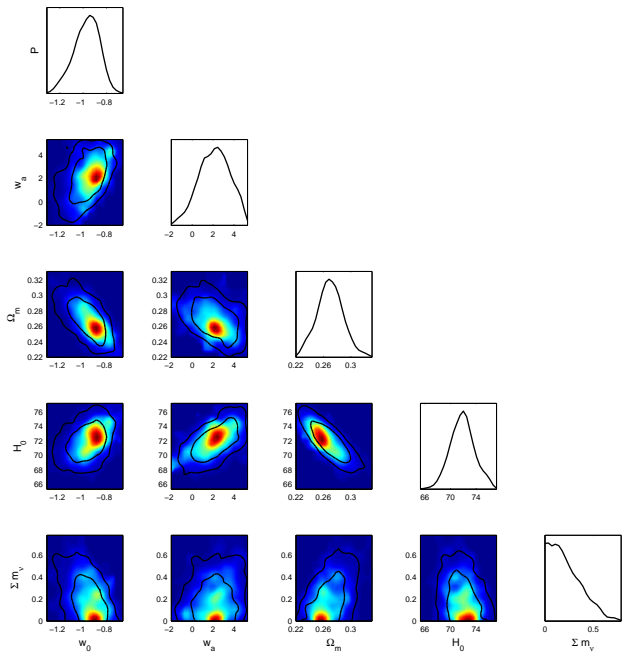}
\caption{Constraints on the cosmological parameters $w_0$, $w_a$,
$\Omega_m$, $H_0$ and $\sum m_{\nu}$ in the dynamical dark energy
scenario (\ref{MZ}).\label{fig2}}
\end{center}
\end{figure}

In Fig.~\ref{fig2}, we present the two dimensional cross correlation
and one dimensional probability distribution of $\sum m_{\nu}$,
$w_0$, $w_a$, $\Omega_m$ and $H_0$, from the data combination
WMAP+LRG+SN+$H_0$. From the panels of $\sum m_{\nu}$--$w_0$ and
$\sum m_{\nu}$--$w_a$, we find that there is an anti-correlation
between $w_0$ and $\sum m_{\nu}$, but no significant correlation
between $w_a$ and $\sum m_{\nu}$ is observed.

The global fitting gives the constraint on the neutrino mass in our
dynamical dark energy scenario: $\sum m_{\nu}<0.56$ eV ($95\%$ CL).
So, it is interesting to find that in our dynamical dark energy
scenario (\ref{MZ}) the total mass of neutrinos can be well
constrained by the data combination WMAP+LRG+SN+$H_0$. To make a
comparison, we show the results of the constant $w$ model given by
Komatsu et al. \cite{Komatsu:2010fb}: $\sum m_{\nu}<0.71$ eV ($95\%$
CL) from WMAP+LRG+$H_0$, and $\sum m_{\nu}<0.91$ eV ($95\%$ CL) from
WMAP+BAO+SN (where SN is the Constitution data). In our case, even
though there is one more parameter, $w_a$, the constraint on $\sum
m_{\nu}$ is much tighter. This shows that the data combination we
used, WMAP+LRG+SN+$H_0$, is fairly good at constraining the neutrino
mass. To show this more clearly, we also use our data combination to
make a global fitting calculation for the case of the constant $w$
model, and the constraint result we obtained is $\sum m_{\nu}<0.45$
eV (95\% CL).

How does the neutrino mass, $\sum m_{\nu}$, affect the constraint
results of ($w_0$, $w_a$)? We find that, with the non-zero neutrino
mass, the results are:
$$w_0=-0.972\pm0.110~ {\rm and} ~w_a=2.038\pm1.492~{\rm (68\%~ CL)}.$$
So, still, though the current data are consistent with a
cosmological constant, a large range of values of ($w_0$, $w_a$) are
still allowed by the data. Comparing to the case with a zero
neutrino mass, $w_0=-0.921\pm0.098$ and $w_a=1.905\pm1.468$ ($68\%$
CL), we find that the best-fit value of $w_0$ is shifted by $5.5\%$,
and the best-fit value of $w_a$ is shifted by $7.0\%$; the error in
$w_0$ is amplified by $12.24\%$, and the error in $w_a$ is amplified
by $1.63\%$. Obviously, the reason that the influence on the error
of $w_0$ by the presence of $\sum m_{\nu}$ is much severer than that
of $w_a$ is as the same as the case of the spatial curvature: $w_0$
is much more sensitive to the data than $w_a$.

\section{Conclusion}\label{sec:conclu}

With the 7-yr WMAP temperature and polarization data, the power
spectrum of LRGs derived from SDSS DR7, the Type Ia supernova data
from Union2 sample, and the new measurements of $H_0$ from HST, we
have tested the dynamical dark energy parametrization, $w(z) = w_{0}
+ w_{a}\left(\frac{\ln(2+z)}{1+z}-\ln2\right)$, in the presence of
spatial curvature and massive neutrinos, by using a MCMC global fit
method. For a time-dependent equation of state, one must be careful
about the treatment of perturbations in dark energy when $w$ crosses
$-1$. We used the method provided in
Refs.~\cite{XiaPert,Zhao:2005vj} to treat the dark energy
perturbations consistently in the whole parameter space in our
numerical calculations. The sound speed of dark energy, $c_s^2$, is
fixed in our calculation to be 1; note that the value of $c_s^2$ is
actually insensitive to our final fit results. Our focus was put on
the determinations of the spatial curvature of the observable
universe, $\Omega_k$, and the total mass of neutrinos, $\sum
m_{\nu}$, in our dynamical dark energy scenario, and the influence
of these factors to the constraints on the dark energy parameters,
$w_0$ and $w_a$.

The $95\%$ CL limit on the spatial curvature is
$-0.0153<\Omega_k<0.0167$. Thus, we find that in our dynamical dark
energy model the spatial curvature is well constrained by
WMAP+LRG+SN+$H_0$. In a curved universe, the constraint results of
$w_0$ and $w_a$ are $w_0=-0.922\pm0.123$ and $w_a=1.651\pm1.470$
($68\%$ CL). Comparing to the case in a flat universe,
$w_0=-0.921\pm0.098$ and $w_a=1.905\pm1.468$ ($68\%$ CL), we find
that the error in $w_0$ is amplified by $25.51\%$, and the error in
$w_a$ is almost the same, only amplified by $0.14\%$. Our global fit
gives the constraint on the neutrino mass in our dynamical dark
energy scenario, $\sum m_{\nu}<0.56$ eV ($95\%$ CL). So, we find
that in our dynamical dark energy scenario the total mass of
neutrinos can be well constrained by the data combination
WMAP+LRG+SN+$H_0$. With the non-zero neutrino mass, the constraint
results of $w_0$ and $w_a$ are $w_0=-0.972\pm0.110$ and
$w_a=2.038\pm1.492$ ($68\%$ CL). Comparing to the case with a zero
neutrino mass, $w_0=-0.921\pm0.098$ and $w_a=1.905\pm1.468$ ($68\%$
CL), we find that the error in $w_0$ is amplified by $12.24\%$, and
the error in $w_a$ is amplified by $1.63\%$.

We believe that it is fairly important to use some divergence-free
parametrization to probe the dynamical evolution of dark energy. We
have shown that the form (\ref{MZ}) is a good proposal, and it has
been proven to be very successful in exploring the properties of
dark energy. We suggest that this parametrization should be further
investigated.

\begin{acknowledgments}
We acknowledge the use of the Legacy Archive for Microwave
Background Data Analysis (LAMBDA). Support for LAMBDA is provided by
the NASA Office of Space Science. The calculation is taken  on
Deepcomp7000 of Supercomputing Center, Computer Network Information
Center of Chinese Academy of Sciences. This work was supported in
part by the National Science Foundation of China under Grant Nos.
11175042, 10975032, 11033005 and 10705041, by the 973 program under
Grant No.~2010CB83300, and by the National Ministry of Education of
China under Grant Nos. NCET-09-0276 and N100505001.
\end{acknowledgments}

\end{document}